\documentclass[12pt,a4paper,superscriptaddress,pre,preprint,showpacs]{revtex4}
\usepackage{graphicx}
\usepackage{amsmath,amsfonts} 

\begin{document}

\newcommand{\EQ}{Eq.~}
\newcommand{\EQS}{Eqs.~}
\newcommand{\FIG}{Fig.~}
\newcommand{\FIGS}{Figs.~}
\newcommand{\SEC}{Sec.~}
\newcommand{\SECS}{Secs.~}

\title{Geographical threshold graphs with small-world and scale-free
properties}
\author{Naoki Masuda}
\affiliation{Laboratory for
Mathematical Neuroscience, RIKEN Brain Science Institute, 2-1,
Hirosawa, Wako, Saitama 351-0198, Japan}
\affiliation{Aihara Complexity Modelling Project, ERATO, JST, 3-23-5, Uehara,
Shibuya, Tokyo 151-0064, Japan}
\author{Hiroyoshi Miwa}
\affiliation{Department of Informatics, School of Science and
Technology, Kwansei Gakuin University, 2-1, Gakuen, Sanda, Hyogo
669-1337, Japan}
\author{Norio Konno}
\affiliation{Faculty of Engineering, Yokohama National University,
79-5, Tokiwadai, Hodogaya, Yokohama 240-8501, Japan}
\date{Received 15 September 2004}

\begin{abstract}
Many real networks are equipped with short diameters, high clustering,
and power-law degree distributions. 
With preferential
attachment and network growth,
the model by Barab\'{a}si and Albert
simultaneously reproduces these
properties, and geographical versions of growing networks
have also been analyzed.
However, nongrowing networks with
intrinsic vertex weights often explain these features more
plausibly, since not all networks are really growing.  We propose
a geographical nongrowing 
network model with vertex weights.  Edges are assumed to
form when a pair of vertices are spatially close and/or have large
summed weights.  Our model generalizes a variety of models as
well as the original nongeographical counterpart, such as the unit
disk graph, the Boolean model, and the gravity model, which
appear in the contexts of percolation, wire communication, mechanical
and solid physics, sociology, economy, and marketing.  In appropriate
configurations, our model produces small-world networks with power-law
degree distributions.  We also discuss the relation between
geography, power laws in networks, and power laws in general
quantities serving as vertex weights.
\end{abstract}

\maketitle

\section{Introduction}\label{sec:introduction}

Networks of interacting agents such as humans, computers,
animal species, proteins, and neurons have been investigated vigorously.
They are typically complicated, meaning that
their structures are far from absolutely regular or entirely random.
Two principal quantities characterizing networks are the
average shortest path length $L$ and the clustering coefficient
$C$. The number of edges in the shortest path averaged over all vertex
pairs defines $L$. Most real networks have small $L$, namely
$L\propto \log n$ or even less, where $n$ is the number of vertices.
The local clustering coefficient is the normalized
number of connected triangles containing a specific vertex.  If the
vertex degree is $k$, or there are $k$ edges adjacent to the vertex,
the normalization factor is $k(k-1)/2$. This quantity averaged over
all the vertices defines $C$, and real networks usually have large
$C$.  A small $L$ and a large $C$ cannot be simultaneously realized either
by lattices, trees, or the ordinary random graphs \cite{SW,reviews}.
Then, Watts and Strogatz proposed the small-world networks
that fulfill these requirements at the same time 
\cite{SW}.

Another important observation is that not all but
many real networks have power-law degree distributions $p(k)\propto
k^{-\gamma}$, typically with scaling exponent $2< \gamma < 3$
\cite{reviews}. The small-world networks are short of
the scale-free property. In light of this, Barab\'{a}si and Albert (BA) proposed a
network model that generates
scale-free networks
with $\gamma=3$ \cite{reviews}. Two essential features of the BA model
are (i) network growth realized 
by sequentially adding vertices and edges, and (ii)
preferential attachment, meaning that newly introduced edges are more
prone to be linked to vertices with larger $k$. Since the proposal of the 
BA model, its various extensions and related models, such as the
fitness model and the hierarchically growing models, have been presented.
These models are successful in incorporating more realistic 
aspects of networks including
tunable $\gamma$ and large $C$ that the original BA model
actually lacks
\cite{reviews,Bianconi,Manna,Xulvi,Yook02,Sen03,Ravasz,Barrat}.

To sum up, some BA-type models and hierarchical networks own
large $C$, small $L$, and scale-free $p(k)$.  However, real
scale-free networks are not necessarily growing.  The number of
vertices may not change greatly over time for networks of friends,
companies, interacting proteins, and neurons, to name a few. In view
of this, a class of nongrowing scale-free networks has been studied
in which whether edges are created relies on
interaction of vertices with intrinsic weights
\cite{Caldarelli}. Weights represent the fitness of
vertices to win edges \cite{Caldarelli,Goh_Chung,Toroczkai}
and are interpreted as, for example, capitals, social skills,
activity levels, information contents, concentration or mass of
physical or chemical substances, and the vertex degree itself.
The role of such vertex fitness was
argued in growing models as well \cite{Bianconi}.
To our
surprise, scale-free $p(k)$ emerges even from weight distributions
devoid of power laws \cite{Caldarelli,Boguna,Servedio,Masuda_THRESH}.
As a remark, if an edge exists when the sum of two vertices' weights
exceeds a prescribed threshold, the network is equivalent to the so-called
threshold graph \cite{Golumbic}. This case
eases analytical treatments.

Our focus in this paper is the geographical extension
of the nongrowing scale-free networks, which has been overlooked so far.
Actually, real networks are often
embedded in topological spaces.  Even the Internet, in which
the speed of information transmission is technically independent of
the physical distance, is subject to geographical constraints because of
wiring costs \cite{Xulvi,Yook02,Sen03,Rozenfeld}.  In addition, it is
often advantageous to map nonphysical quantities or networks into
geographical spaces by, for example,
the principal component analysis.
Then, influence of the distance between
graduated traits is questioned.

In fact, the Watts-Strogatz small-world network already addressed this
issue since it is constructed on lattice substrates 
\cite{SW}. Let us denote by
$g(r)$ the probability that two vertices with distance $r$ are
connected. In lattice networks supplied with additional edges,
where $g(r)\propto r^{-\delta}$,
generated networks have small $L$ when $\delta$ is smaller than a
critical value
\cite{Sen02,Biskup03,Kleinberg_Franceschetti}.  Otherwise, global
connections are too scarce to elicit the small-world property. The same is
true for growing scale-free networks. Although the BA model is
irrelevant to embedding spaces, which is actually the main cause for
small $C$, it has been extended to incorporate
underlying geographical spaces and $g(r)=r^{-\delta}$.  Then, a
transition from the scale-free to nonscale-free regime as well as
one from small $L$ to large $L$ occurs at a certain $\delta$
\cite{Manna,Xulvi,Yook02,Sen03}.

Although $g(r)$ plays a key role in determining the network structure,
characterization of $g(r)$ in real-world
networks still seems controversial.  In applications such as the
Internet routing \cite{Waxman} and neural networks \cite{Amari_etc},
$g(r)$ decaying exponentially or in a Gaussian manner
is
commonly used.  Exponential decays are also inferred from biological
neural networks \cite{Hellwig}.  However, many other data are in favor of
$g(r)\propto r^{-\delta}$.  For instance, a recent extensive
analysis of the Internet concludes $g(r)\propto r^{-1}$
\cite{Yook02}. Power laws also hold for macroscopic connectivity of
brain regions identified by correlated activities [$g(r)\propto
r^{-2}$] \cite{Eguiluz03} and for microscopic neural networks
\cite{Karbowski}.  In social sciences, evaluating $g(r)$ seems more
difficult because of presumably larger noises. Accordingly, both
power-law and exponential forms of $g(r)$ have been inferred,
sometimes even from identical data \cite{Haynes,Hutchinson}.
In the face of the ambiguity of available data, 
it is worthwhile to examine models to see how various types of
$g(r)$ affect network properties to help
interpret real data.

In the context of nongrowing geographical networks,
there is an algorithm that generates $p(k)=k^{-\gamma}$
with a prescribed $\gamma$ \cite{Rozenfeld}.
However, investigations of nongrowing geographical networks are largely
missing, particularly when interaction of vertices, which 
is not considered in  \cite{Rozenfeld}, takes place.
We examine a geographical 
threshold network model with various configurations.
In \SEC\ref{sec:caldarelli}, we review the nongeographical threshold
model with vertex weights. In \SEC\ref{sec:model}, 
we introduce the geographically extended model and analyze some
practical cases, including
the unit disk graph and the gravity model. 
Section~\ref{sec:discussion} is devoted
to discussing our model in the context of network search problems and
real data.

\section{Nongeographical threshold network model}\label{sec:caldarelli}

Before taking geography into account, we briefly summarize
the ordinary 
threshold network model, which constitutes
a subclass of networks with intrinsic vertex weights
\cite{Caldarelli,Boguna,Servedio,Masuda_THRESH}.

We prepare $n$ vertices denoted by $v_i$ ($1\le i\le n$), each 
of which carries
a weight variable $w_i\in {\mathbb R}$
randomly and independently distributed as specified by the
density function $f(w)$.
As mentioned in
\SEC\ref{sec:introduction},
$w_i$ quantifies the propensity
for $v_i$ to gain edges.
Let
\begin{equation}
F(w) = \int^w_{-\infty} f(w^{\prime}) dw^{\prime}
\end{equation}
be the cumulative distribution function.
We explain with additive weights since 
multiplicative weights are transformed into
additive weights by taking the logarithm.
In the nongeographical threshold model, an edge exists
between $v_i$ and $v_j$ ($i\neq j$) if and only if 
$w_i+w_j\ge \theta$. When $n$ is sufficiently large, the weight $w$
uniquely determines the vertex degree $k$ by
\begin{eqnarray}
k &=& n \left[ 1 - F\left(\theta - w\right) \right].
\label{eq:k(w)}
\end{eqnarray}
Using \EQ(\ref{eq:k(w)}), the degree distribution $p(k)$
($0\le k< n$) is written as
\begin{eqnarray}
p(k) &=& f(w)\frac{dw}{dk}\nonumber\\
&=& \frac{f\left[\theta-F^{-1}\left( 1 - \frac{k}{n}\right)\right]}
{n f\left[ F^{-1}\left(1 - \frac{k}{n}\right)\right]}.
\label{eq:p(k)}
\end{eqnarray}
Because the model is simple, $L$, $C$, and the correlation
between the degrees of adjacent vertices can be
analytically derived as well \cite{Boguna,Masuda_THRESH}. The small-world
properties characterized by a large $C$ and a small $L$ are fulfilled
for a wide choice of $f(w)$.  More microscopically, vertices with
small degrees have $C(k)$ near 1 and form the peripheral part of the
network. It is connected to the cliquish
core with larger $k$ and smaller $C(k)$.  Strictly speaking, the core
consists of the vertices with $w\ge \theta/2$, and the
separability of this kind is known in the graph theory
\cite{Golumbic}. A similar
separability is also mentioned in other literature \cite{Goh_Chung}.

The degree distribution depends on $f(w)$. An easily solvable example is
the exponential weight distribution given by
\begin{equation}
f(w) = \lambda e^{-\lambda w} \quad
(0\le w).
\label{eq:f(w)_exp}
\end{equation}
We set $\theta > 0$ because otherwise 
the network becomes the complete
graph. Although $f(w)$ in \EQ(\ref{eq:f(w)_exp}) is not
reminiscent of the power law, substitution of
\EQ(\ref{eq:f(w)_exp}) into \EQ(\ref{eq:p(k)}) yields
$p(k)\propto k^{-2}$ \cite{Caldarelli}.
It follows that $C(k)\propto k^{-2}$ and
$\overline{k}(k)\propto k^{-1}$, where
$\overline{k}(k)$ is a measure of degree correlation, namely,
the average degree of the neighbors of
a reference vertex with degree $k$  \cite{Boguna,Masuda_THRESH}. 
The same scaling law is also maintained
for the logistic distribution $f(w) = \beta
e^{-\beta w} / \left( 1 + e^{-\beta w}\right)^2$, which is
just a slight modification of \EQ(\ref{eq:f(w)_exp})
\cite{Masuda_THRESH}.
Another major class of $f(w)$ is the Pareto distribution defined by
\begin{equation}
f(w) = \frac{a}{w_0}\left( \frac{w_0}{w}\right)^{a+1}
\quad (w\ge w_0),
\label{eq:f(w)_pareto}
\end{equation}
where $a, w_0>0$. Equation~(\ref{eq:f(w)_pareto}) leads to
$p(k)\propto k^{-\gamma}$ with $\gamma=(a+1)/a>1$, $C(k)\propto
k^{-1}$, and $\overline{k}(k)\propto
 k^{-1}$ \cite{Masuda_THRESH}. Particularly,
$C(k)\propto k^{-1}$ is more consistent with real data \cite{Ravasz}
compared with $C(k)\propto k^{-2}$ derived from
\EQ(\ref{eq:f(w)_exp}).  The asymptotics is the same for the Cauchy
distribution $f(w) = 1/[\pi (1+w^2)]\;$ $(w\in {\mathbb R})$, which is
devoid of the lower bound of $w$.
The inverse problem to determine $f(w)$
from $p(k)$ has also been addressed \cite{Servedio}.

A crux of the threshold model is that scale-free $p(k)$ results from a
wide class of $f(w)$. Analytical and numerical evidence
indicates that $\gamma=2$ is the baseline scaling exponent of the
threshold model, which contrasts with $\gamma=3$ for the
BA model \cite{Masuda_THRESH}. Since the effect of
a lower bound of $w$ seems marginal,
we mainly use the exponential and Pareto
distributions for the geographically extended model.

\section{Geographical threshold network model}\label{sec:model}

To generalize the model introduced in \SEC\ref{sec:caldarelli} in the
geographical case, we assume that vertices are uniformly and
independently distributed with density $\rho$ in a $d$-dimensional
Euclidean space whose coordinates are denoted by $x_1$, $x_2$,
$\ldots$, $x_d$.  Then a pair of vertices with weights $w$,
$w^{\prime}$, and Euclidean distance $r$
are connected if and only if
\begin{equation}
(w+w^{\prime}) h(r)\ge \theta,
\label{eq:rule_h}
\end{equation} 
where $h(r)$ is assumed to decrease in $r$, although
$h(r)$ increasing in $r$ has also been considered in other models
\cite{Manna,Sen03,Sen02}. As a special case,
\EQ(\ref{eq:rule_h}) with $h(r)\propto r^{-1}$ is equivalent
to the Boolean model \cite{Meester_Roy}.

Based on \EQ(\ref{eq:rule_h}), two vertices with weights 
$w$ and $w^{\prime}$ are adjacent if
\begin{equation}
r\le h^{-1}\left( \frac{\theta}{w+w^{\prime}}\right).
\end{equation}
For a vertex with weight $w$, the degree $k$ is represented by 
\begin{equation}
k = \int f\left(w^{\prime}\right) dw^{\prime}
\left[\mbox{number of vertices in a ball of radius }=
h^{-1}\left( \frac{\theta}{w+w^{\prime}}\right)\right].
\label{eq:rule_k}
\end{equation}
This recovers a general formulation \cite{Caldarelli,Boguna}, in
which $k$ is calculated from the joint probability
as a function of $w$ and $w^{\prime}$
that a pair of vertices are connected.
Combination of \EQ(\ref{eq:rule_k}) and $f(w)$ provides
$p(k)$. If we take an average over $w$ but not
over $r$, we obtain $g(r)$. Although $g(r)$ decreases in $r$ if
$h(r)$ does, it generally differs from $h(r)$.

\subsection{Unit disk graph}\label{sub:unit}

If $f(w)=\delta(w_0)$,
where $\delta$ is the delta function, two vertices are
adjacent when $2w_0 h(r)\ge \theta$. Since $h(r)$
decreases in $r$, this condition is equivalent to $r\le r_0$ where
$2w_0 h(r_0) = \theta$. Accordingly,
\begin{equation}
g(r) = \left\{ \begin{array}{ll}
1 & (r\le r_0),\\
0 & (r > r_0),
\end{array}\right.
\label{eq:unit_upper}
\end{equation}
and the generated network is the unit disk graph, which
is applied to modeling broadcast and sensor networks
\cite{Golumbic,Clark_Fishkin}.
If $f(w)$ has a finite support, the
network resembles the unit disk graph in the sense that
there exists an upper limit $r=r_0$ only below which
$g(r)>0$. With this case included,
long-range edges are entirely
prevented, and the network has $L\propto n^{1/d}$, spoiling the
small-world property.  However, if we allow $g(r)=p$ ($r>r_0$) with
$0<p \cong n^{-1} \ll 1$, we have
a type of the Watts-Strogatz small-world
networks with small $L$ \cite{reviews}.
Even so, $p(k)$ is essentially homogeneous.
To introduce the scale-free property, we need to use
more inhomogeneous vertex weights.

\subsection{Exponential weight distribution with $h(r)\propto 
r^{-\beta}$}\label{sub:exponential_power}

Let us consider the exponential weight distribution
given in \EQ(\ref{eq:f(w)_exp}) and set
\begin{equation}
h(r) = r^{-\beta},
\label{eq:r_power}
\end{equation}
where $\beta\ge 0$.
This case generalizes
the nongeographical model explained in \SEC\ref{sec:caldarelli},
which corresponds to $\beta=0$.
For a larger $\beta$, geographical effects are more manifested.
As a function of the weight, the degree
is calculated as follows:
\begin{eqnarray}
k(w) &=& \int^{\infty}_0 f(w^{\prime}) dw^{\prime}\;
\int_{(w+w^{\prime})/r^{\beta}\ge \theta}  \rho \; dx_1 \ldots
dx_d\nonumber\\
&=& \rho \int^{\infty}_0 \lambda e^{-\lambda w^{\prime}}
\pi^{d/2}\Gamma\left(\frac{d}{2}+1\right)
\left( \frac{w+w^{\prime}}{\theta} \right)^{d/\beta}
\; dw^{\prime}
\nonumber\\
&=& c_1
 e^{\lambda w}
\Gamma\left( \frac{d}{\beta}+1, \lambda w \right),
\label{eq:k_exp_power}
\end{eqnarray}
where
\begin{equation}
\Gamma(\alpha, x) \equiv \int^{\infty}_x t^{\alpha-1}e^{-t} dt.
\label{eq:incomplete_gamma}
\end{equation}
is the incomplete Gamma function,
\begin{equation}
c_1\equiv \frac{\rho \pi^{d/2}}{(\theta\lambda)^{d/\beta}}
\Gamma\left(\frac{d}{2}+1\right),
\end{equation}
and 
$\Gamma(\alpha) \equiv \Gamma(\alpha,0)$ is the ordinary Gamma function.
To obtain $p(k)$ from
$k(w)$, we just need to eliminate $w$ from 
\EQ(\ref{eq:k_exp_power})
as we have done in \EQ(\ref{eq:p(k)}).

However, an analytical form of $p(k)$ corresponding to
\EQ(\ref{eq:k_exp_power}) is unavailable due to
the incomplete Gamma function. Accordingly,
let us deal with some
special cases.  By integrating
\EQ(\ref{eq:incomplete_gamma}) by parts, we derive
\begin{equation}
\Gamma(\alpha, x) = (\alpha-1)!\; e^{-x}
\sum^{\alpha-1}_{\alpha^{\prime}=0}
\frac{x^{\alpha^{\prime}}}{\alpha^{\prime}!}
\quad (\alpha\in {\mathbb Z}),
\label{eq:incomplete_gamma_special}
\end{equation}
where ${\mathbb Z}$ is the set of integers.
In the limit $\beta\to 0$,
\EQ(\ref{eq:incomplete_gamma_special}) implies
\begin{equation}
\lim_{\beta\to 0} \frac{\Gamma\left(\frac{d}{\beta}+1,\lambda w
  \right)}
{\left(\frac{d}{\beta}\right)!} = 1.
\label{eq:limit_beta0}
\end{equation}
Actually, $k$ explodes as $\beta\to 0$
because \EQ(\ref{eq:limit_beta0}) means
$\lim_{\beta\to 0}
\Gamma\left(\left(d/\beta\right)+1,\lambda w \right) = \infty$,
reflecting the density notation
of the vertex distribution.
Putting aside this nonessential point,
$\Gamma\left(\left(d/\beta\right)+1,\lambda w \right)$ is asymptotically
independent of $w$, and we have
\begin{equation}
k(w)\propto e^{\lambda w}
\end{equation}
and
\begin{equation}
p(k)\propto e^{-2\lambda w}\propto k^{-2},
\label{eq:p(k)_beta0}
\end{equation}
which reproduces the results for the nongeographical counterpart
\cite{Caldarelli,Boguna,Masuda_THRESH}.
For a sufficiently small 
$\beta$, \EQ(\ref{eq:limit_beta0})
effectively approximates the incomplete Gamma function.
Consequently, scale-free $p(k)$ with $\gamma= 2$
or a slightly larger $\gamma$ is almost preserved.

When $\beta=d$, we obtain
\begin{equation}
k(w) = c_1 e^{\lambda w}
\Gamma\left( 2, \lambda w \right) = c_1 (1+\lambda w)
\label{eq:k_exp_power_example2}
\end{equation}
and
\begin{equation}
p(k) = \frac{\lambda e^{-\lambda w}}{c_1 \lambda}
= \frac{\exp\left(1-\frac{k}{c_1}\right)}{c_1}.
\label{eq:beta=d}
\end{equation}
The degree distribution now has an exponential tail, and hubs
are less likely compared with \EQ(\ref{eq:p(k)_beta0}).
Another special case with $\beta=d/2$ leads to
\begin{equation}
k(w) = c_1 \left(2+2\lambda w + \lambda^2 w^2\right)
\label{eq:k_exp_power_example3}
\end{equation}
and
\begin{equation}
p(k) = \frac{\lambda e^{-\lambda w}}{2c_1 (\lambda + \lambda^2
  w)} = \frac{\exp\left(1-\sqrt{\frac{k}{c_1}-1}\right)}{2c_1^{3/2}\sqrt{k-c_1}}.
\label{eq:beta=d/2}
\end{equation}
Equation~(\ref{eq:beta=d/2}) is a stretched exponential distribution
with a minor modification factor $k^{-1/2}$, and $p(k)$ decays more
slowly than in \EQ(\ref{eq:beta=d}) naturally because $\beta=d/2 < d$.

Similarly, $k$ and $w$ are connected by a power law relation if
$\beta > 0$. Then, $p(k)$ is a type of stretched exponential.  In
geographical preferential attachment models, the crossover from a
power-law tail to a stretched exponential tail occurs at a finite
value of the control parameter similar to $\beta$
\cite{Manna,Xulvi,Yook02,Sen03}.  We could say that, in our model, the
same transition happens at $\beta=0$. However, the gist is that for a
sufficiently small $\beta$, $p(k)$ is practically indiscernible from
the scale-free distribution.

\begin{figure}
\begin{center}
\includegraphics[height=2.25in,width=2.25in]{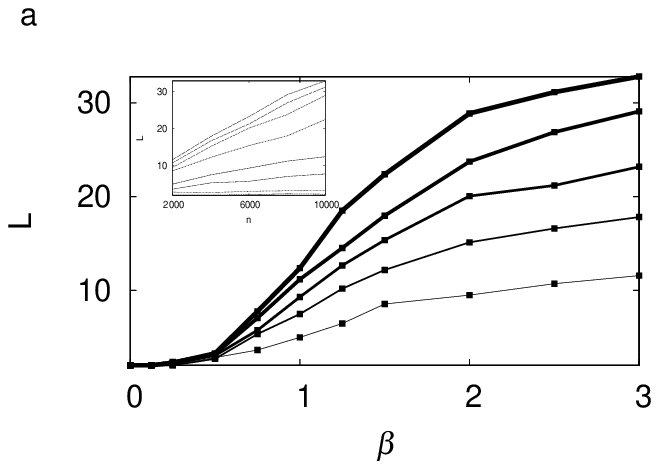}
\includegraphics[height=2.25in,width=2.25in]{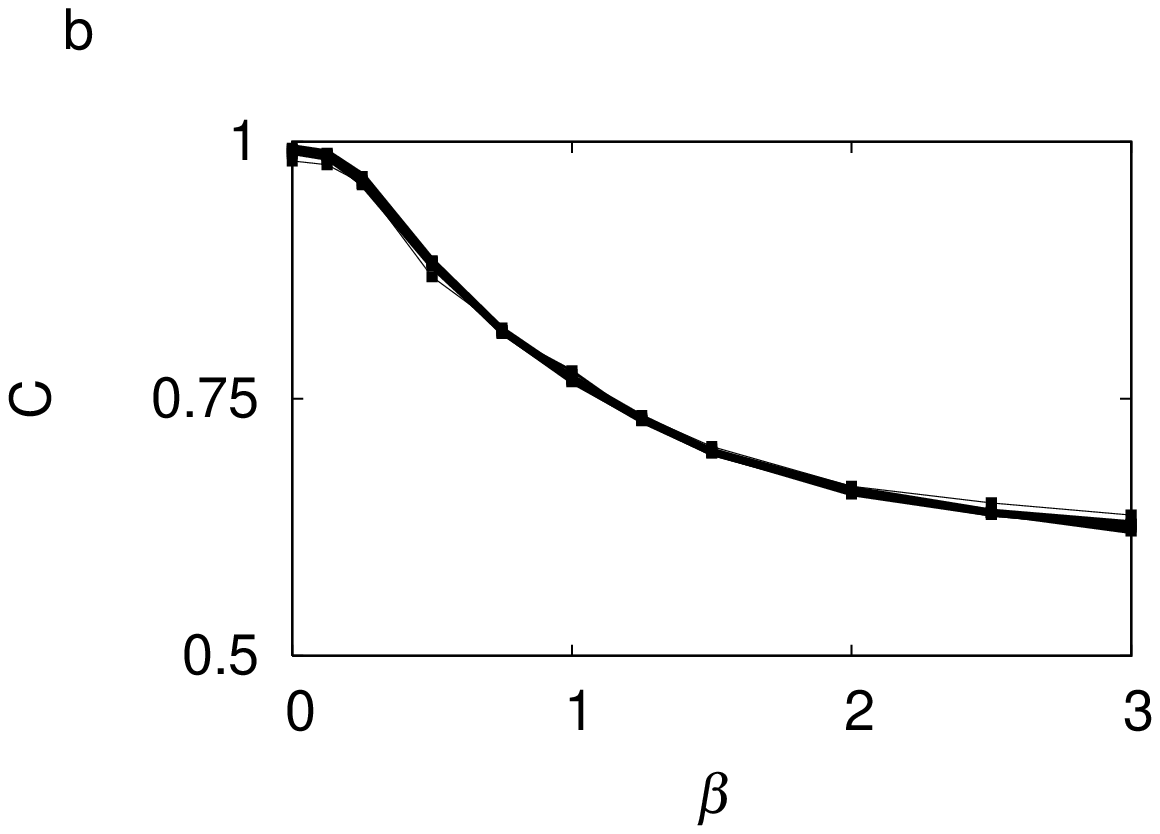}
\includegraphics[height=2.25in,width=2.75in]{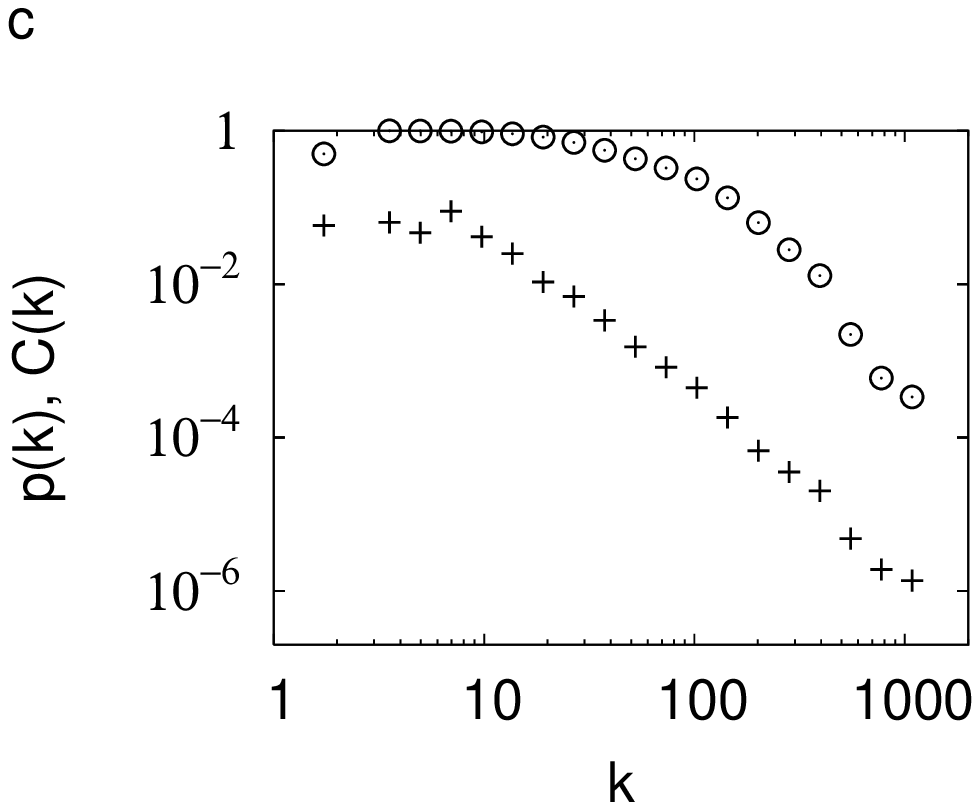}
\includegraphics[height=2.25in,width=2.75in]{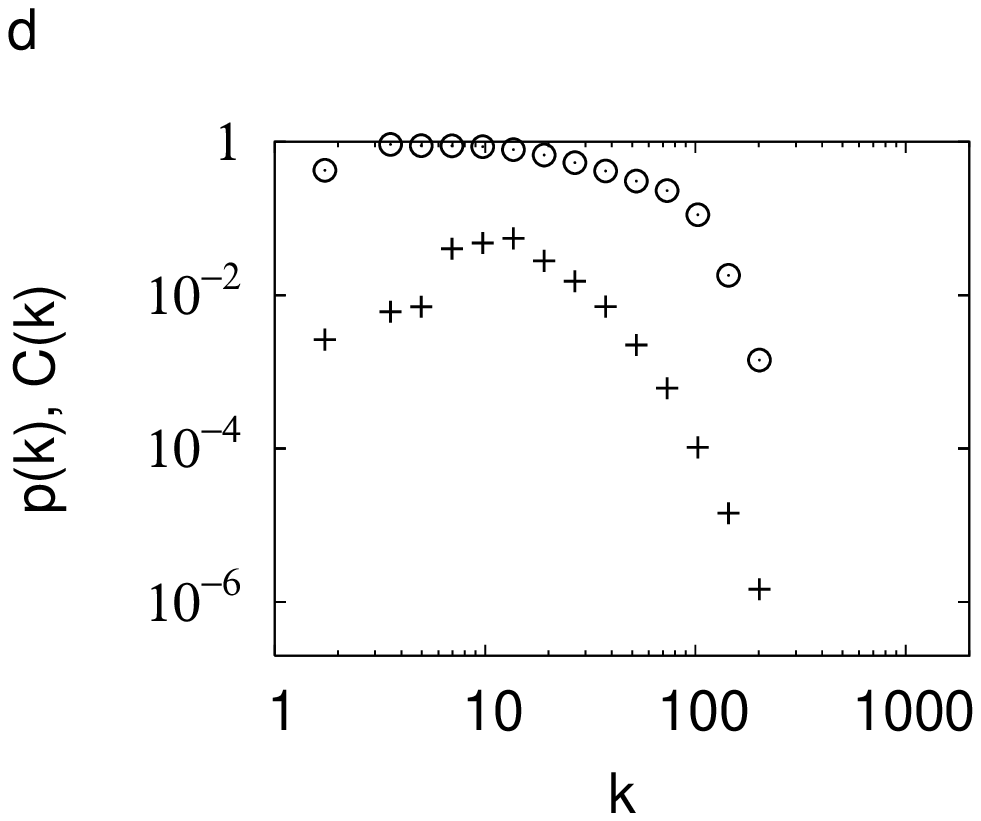}
\includegraphics[height=2.25in,width=2.75in]{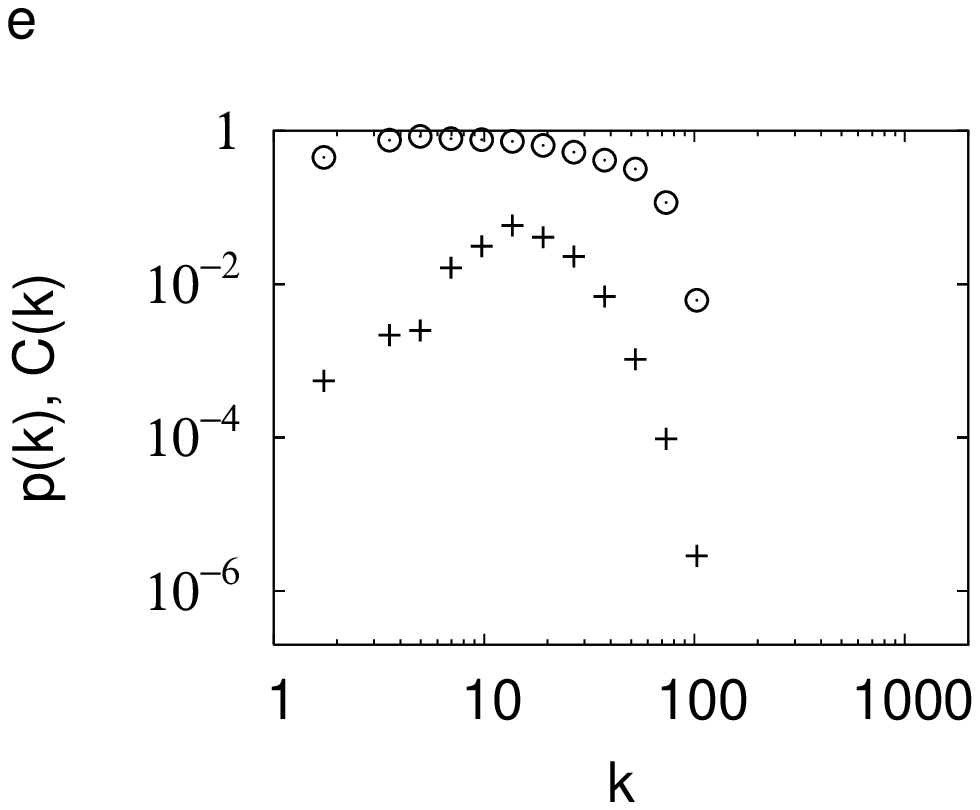}
\caption{}
\label{fig:exp_add}
\end{center}
\end{figure}

Since it seems difficult to analytically calculate other network
characteristics such as $L$ and $C$, we resort to
numerics.
We uniformly scatter $n=10000$ vertices in a two-dimensional square
lattice with side length $100$ and periodic
boundary conditions. Because more
edges obviously means smaller $L$, the mean degree denoted by $\left< k
\right>$ is kept at 20. The analytic expression for $\left< k
\right>$ is available only when $\beta=0$ as follows
\cite{Boguna,Masuda_THRESH}:
\begin{equation}
\left< k \right> 
= e^{-\lambda\theta} \left( \rho l^d + \lambda\theta \right),
\end{equation}
where $l$ is the side length of the area.
Therefore,
we manually modulate $\theta$ to preserve $\left< k \right>$ as we
vary $\beta$.
Excluding isolated components, which actually consist
of just a few vertices,
we show a dependence of $L$ and $C$ on $\beta$ in
\FIGS\ref{fig:exp_add}(a) and \ref{fig:exp_add}(b), respectively.
Although the main simulations are done
with $n=10000$ (thickest lines),
we also show results for $n=2000$ (thinnest lines), $4000$, $6000$,
and $8000$. The inset of \FIG~\ref{fig:exp_add}(a) shows the dependence of 
$L$ on $n$, with
upper lines corresponding to larger values of $\beta$.
Figure~\ref{fig:exp_add}(a) shows that $L$ is insensitive to
$n$ only when $\beta < 0.5$.
We expect that $L\propto \log n$ approximately holds
in this regime. On the other hand, we expect 
$L\propto n^{1/d}$ or similar scaling for larger $\beta$.
As $\beta$ increases,
$C$ decrease to some extent but not too much
to spoil the clustering property [\FIG\ref{fig:exp_add}(b)].
Figures~\ref{fig:exp_add}(c), 
\ref{fig:exp_add}(d), and \ref{fig:exp_add}(e)
show $p(k)$ (crosses) and $C(k)$ (circles)
for $\beta=0.5$, $\beta=1.5$, and $\beta=2.5$, respectively.
As expected, small $\beta$ yields a long tail indicative of
the power law [\FIG\ref{fig:exp_add}(c)].
In contrast, \FIG\ref{fig:exp_add}(e) shows that
$p(k)$ decays much faster when $\beta$ is larger. 
Consequently, networks generated by 
sufficiently small $\beta$ are
endowed with
the scale-free and small-world properties simultaneously in
a geographical context,
which extrapolates the nongeographical results with
$\beta=0$. 
In regard to the vertex-wise clustering coefficients, $C(k)\propto
k^{-2}$ holds when $\beta=0$
\cite{Boguna,Masuda_THRESH}. The numerical results in
\FIGS\ref{fig:exp_add}(c), \ref{fig:exp_add}(d), and \ref{fig:exp_add}(e)
(circles) support $C(k)\propto k^{-2}$ except that
vertices with small $C(k)$ are more scarce for larger $\beta$.

The probability
$g(r)$ that two vertices with distance
$r$ are adjacent becomes
\begin{eqnarray}
g(r) &=& \int^{\infty}_0
 \lambda e^{-\lambda w} dw
\int_{(w+w^{\prime})/r^{\beta}\ge \theta} 
 \lambda e^{-\lambda w^{\prime}} dw^{\prime}\nonumber\\
&=& \int^{\theta r^{\beta}}_0 \lambda e^{-\lambda w} dw
\int^{\infty}_{\theta r^{\beta}-w} 
 \lambda e^{-\lambda w^{\prime}} dw^{\prime}
+ \int^{\infty}_{\theta r^{\beta}} \lambda e^{-\lambda w} dw
\nonumber\\
&=& e^{-\lambda\theta r^{\beta}} (\lambda \theta r^{\beta}+1),
\label{eq:g(r)_exp}
\end{eqnarray}
indicating a stretched
exponential decay in $r$ unless $\beta=0$.  Particularly, the main decay rates for $\beta = 1$
and $\beta=2$,
respectively, correspond to the standard exponential and the Gaussian,
which are
widely used in applications \cite{Waxman,Amari_etc}. As a general
remark,
$g(r)$ does not coincide with $h(r)\propto r^{-\beta}$ even asymptotically.

The loss of the small-world property for large $\beta$ seems to stem
from the (stretched) exponential decay of $g(r)$. In addition,
$g(r)$ derived here qualitatively
disagrees with many real data
\cite{Yook02,Eguiluz03,Karbowski}. As a result,
exponential types of $g(r)$ and the Gaussian $g(r)$
may be far from universal. This
is a striking caveat to many fields,
such as neuroscience, social dynamics,
and epidemics, which
conventionally assume
geographical networks with exponentially decaying or Gaussian $g(r)$.
We do not explore consequences of $h(r)$ that
decays faster than $h(r)\propto r^{-\beta}$, since such an $h(r)$ must
yield even larger $L$. On the other hand, $h(r)$
with slower decays, or $h(r)\propto \left(\log r\right)^{-1}$, is
examined in \SEC\ref{sec:gravity}.

\subsection{Power-law weight distribution with $h(r)\propto r^{-\beta}$}\label{sub:pareto_power}

Quantities that can serve as vertex weights, such as the city and firm
sizes \cite{Zipf47,Zipflaw,Rosen,Axtell}, number of pages in a
website \cite{Huberman}, land prices \cite{Kaizoji_Andersson},
incomes \cite{Levy_Okuyama}, importance of airports \cite{Barrat}, 
and importance of academic authors
\cite{Barrat}, are often distributed according to power
laws.  The history of these power laws 
is much longer, dating back to 
the Pareto and Zipf laws, 
than those recently
found for networks \cite{reviews}.  The simplest way to associate the
power laws of networks with those of vertex weights is simply to
interpret the vertex degree as the weight. However, $w$ and $k$ are
generally nonidentical \cite{Barrat,Masuda_THRESH}.

Let $f(w)$ be the Pareto distribution given in 
\EQ(\ref{eq:f(w)_pareto}).
With the interaction strength decaying algebraically
[\EQ(\ref{eq:r_power})], we have
\begin{eqnarray}
k(w) &=& \rho \int^{\infty}_{w_0}
\frac{a}{w_0}\left(\frac{w_0}{w^{\prime}}
\right)^{a+1} \pi^{d/2}
\Gamma\left(\frac{d}{2}+1\right)
\left( \frac{w+w^{\prime}}{\theta}\right)^{d/\beta}
dw^{\prime} \nonumber\\
&=& \frac{a w_0^a \rho \pi^{d/2}}{\theta^{d/\beta}}
\Gamma\left(\frac{d}{2}+1 \right) w^{(d/\beta)-a}
\int^{\infty}_{w_0/w} \frac{(1+x)^{d/\beta}}{x^{a+1}} dx
\quad (w\ge w_0).
\label{eq:k(w)_pareto_power}
\end{eqnarray}
Convergence of $k(w)$ necessitates $-a+d/\beta<0$.
In the limit $w\to\infty$, it holds that
\begin{equation}
\int^{\infty}_{w_0/w} \frac{(1+x)^{d/\beta}}{x^{a+1}} dx
 \propto
\int^{\infty}_{w_0/w} \frac{1}{x^{a+1}} dx
\propto \left(\frac{w}{w_0}\right)^a.
\end{equation}    
Therefore,
\begin{equation}
k(w)\propto w^{d/\beta}
\end{equation}
and
\begin{equation}
p(k)\propto \frac{\frac{a}{w_0}\left(\frac{w_0}{w}\right)^{a+1}}
{\frac{d}{\beta}w^{(d/\beta)-1}}\propto k^{-1-(a\beta/d)}.
\label{eq:p(k)_add_pareto}
\end{equation}
In contrast to the stretched exponential scenario clarified in
\SEC\ref{sub:exponential_power},
the power-law weight distribution
produces scale-free $p(k)=k^{-\gamma}$ with
$\gamma= 1+(a\beta/d)$.

For $r$ large enough to satisfy $\theta r^{\beta}\ge 2 w_0$,
\begin{eqnarray}
g(r) &=& \int^{\infty}_{\theta r^{\beta}-w_0}
\frac{a}{w_0}\left(\frac{w_0}{w^{\prime}}
\right)^{a+1} dw^{\prime}
+ \int^{\theta r^{\beta}-w_0}_{w_0}
\frac{a}{w_0}\left(\frac{w_0}{w^{\prime}}
\right)^{a+1}
\left( \frac{w_0}{\theta r^{\beta}-w^{\prime}}\right)^a dw^{\prime}
\nonumber\\
&=& 
\left( \frac{w_0}{\theta r^{\beta}-w_0}\right)^a
+ \int^{b-1}_1 x^{-a-1}(b-x)^{-a} dx,
\label{eq:g(r)_add_pareto}
\end{eqnarray}
where $b\equiv \theta r^{\beta}/w_0$.
To show that the integral in \EQ(\ref{eq:g(r)_add_pareto})
tends to be proportional to $r^{-a\beta}$ as
$r\to\infty$, let us evaluate
$b^a \int^{b-1}_1 A(x) dx$, where 
$A(x) \equiv x^{-a-1}(b-x)^{-a}$.
First, we obtain
\begin{eqnarray}
\liminf_{b\to\infty}
b^a\int^{b-1}_1 A(x) dx&\ge&
\lim_{b\to\infty}\left(\frac{b}{b-1}\right)^a
\int^{b-1}_1 x^{-a-1} dx\nonumber\\
&=& \frac{1}{a} \lim_{b\to\infty}\left[ 1-\left(b-1\right)^{-a}\right]
= \frac{1}{a}.
\label{eq:g(r)_liminf}
\end{eqnarray}
To bound the integral from the above in the limit $b\to\infty$,
let us assume $b>4$. Noting that 
$A(x)$ takes the minimum at 
$x=(a+1)b/(2a+1)$ and that $d^2 A(x)/dx^2>0$, we derive
\begin{eqnarray}
&& \limsup_{b\to\infty} b^a\int^{b-1}_1 A(x) dx\nonumber\\
&\le& \lim_{b\to\infty}\left\{
\int^{(b+2)/3}_1 \left[ \frac{3b}{2(b-1)}\right]^a x^{-a-1} dx
+ \frac{b^a}{2}\left[ A\left(\frac{b+2}{3}\right)
+A\left(\frac{(a+1)b}{2a+1}\right)\right]
\left[\frac{(a+1)b}{2a+1}-\frac{b+2}{3} \right]\right. \nonumber\\
&&\left. + \frac{b^a}{2}\left[A\left(\frac{(a+1)b}{2a+1}\right)
+A\left(b-1\right)\right] 
\left[ \left(b-1\right)-\frac{(a+1)b}{2a+1}\right] \right\} 
\nonumber\\
&=& \lim_{b\to\infty}\left\{
\left[ \frac{3b}{2(b-1)}\right]^a \frac{1}{a}
\left[ 1-\left(\frac{3}{b+2}\right)^a\right]
\right. \nonumber\\
&&\left. + \frac{1}{2}\left(\frac{b}{b-1}\right)^a
\left[\frac{ab-2a-1}{(2a+1)(b-1)} \right] + O\left(b^{-a}\right)
\right\} \nonumber\\
&=& \frac{1}{a}\left(\frac{3}{2}\right)^a + \frac{a}{2(2a+1)} < \infty.
\label{eq:g(r)_limsup}
\end{eqnarray}
Equations~(\ref{eq:g(r)_liminf}) and (\ref{eq:g(r)_limsup})
allow us to conclude an algebraic decay
$g(r)\propto r^{-a\beta}$ in
contrast to \EQ(\ref{eq:g(r)_exp}).
Discussion on network structure
is postponed to \SEC\ref{sec:gravity}, where we
analyze the gravity model, which ends up with
the same asymptotic behavior of $p(k)$ and $g(r)$.

\subsection{Gravity model with Pareto $f(w)$}\label{sec:gravity}

As shown in \SEC\ref{sub:exponential_power}, given the exponentially
distributed $w$, $h(r)\propto r^{-\beta}$ with a sufficiently small
$\beta$ yields more or less desired network properties. More
rapidly decaying $h(r)$ makes $g(r)$ decrease too
fast to elicit small $L$.  How about $h(r)$
that decays more slowly?  To address this issue, we apply
$h(r)\propto (\log r)^{-1}$. Since $\log r$ can be negative, let us
rewrite \EQ(\ref{eq:rule_h}) as
\begin{equation}
w + w^{\prime}\ge \theta \log r.
\label{eq:rule_h_log}
\end{equation}
Equation~(\ref{eq:rule_h_log}) is equivalent to
\begin{equation}
e^w e^{w^{\prime}}\ge r^\theta.
\label{eq:rule_h_log_exp}
\end{equation}
Since edge formation is suppressed by
increasing either $\beta$ or $\theta$,  
let us reinterpret $\theta$ in \EQ(\ref{eq:rule_h_log_exp}) as $\beta$,
which does not 
essentially change the model.
Further rescaling of the parameters
by $W\equiv e^w$, $W\equiv e^{w^{\prime}}$, and
$R = \theta^{-1/\beta} r$ transforms
\EQ(\ref{eq:rule_h_log_exp}) into
\begin{equation}
\frac{W W^{\prime}}{R^{\beta}}\ge \theta.
\label{eq:gravity_W}
\end{equation}
This is the gravity model often used in physics, sociology, economics,
and marketing \cite{Haynes,Zipf47,Long_Howrey,Fotheringham,Morrill}.
The gravity model is suitable in describing interaction of particles
in geographical spaces when the physical gravity ($\beta=2$) or
similar mass interaction based on, for example, populations or chemical
substances, is active. In the sociological context, the original model
stipulates $\beta=1$ \cite{Zipf47}, but $\beta$ ranging from 0.2 to
2.7 have been inferred from later real data
\cite{Haynes,Hutchinson,Long_Howrey,Fotheringham,Morrill,Latane}.

The original gravity model is geographical but 
neglects weight distributions. On the other hand,
multiplicatively interacting weights with
power-law $f(w)$ are used to generate solvable
scale-free networks, but they ignore geography
\cite{Goh_Chung}. We are interested in 
combined effects of geography and dispersed vertex weights.
The transformation from
\EQ(\ref{eq:rule_h_log}) to \EQ(\ref{eq:gravity_W}) also rescales
$f(w)$ unless it is the delta function.  When the weights in
\EQ(\ref{eq:rule_h_log}) follows the exponential distribution
given in \EQ(\ref{eq:f(w)_exp}), the
density $\overline{f}(W)$ of the weights in
\EQ(\ref{eq:gravity_W}) becomes
\begin{equation} 
\overline{f}(W) = f(w)\frac{dw}{dW}
= \lambda \left(\frac{1}{W}\right)^{\lambda+1},
\end{equation}
namely the Pareto distribution
with $a=\lambda$ and $w_0=1$.
Although we have started with $h(r)\propto (\log r)^{-1}$ and additive
weights, we switch to the gravity-model notation for
convenience.  Now we rewrite \EQ(\ref{eq:gravity_W}) as
\begin{equation}
\frac{w w^{\prime}}{r^{\beta}} \ge \theta
\label{eq:rule_gravity}
\end{equation}
and investigate the network structure
when $f(w)$ is the Pareto distribution.

Before moving on to the Pareto $f(w)$,
let us note that $f(w)$ with a finite support
only allows
local interaction, as explained in \SEC\ref{sub:unit}.
Then the gravity model yields
$L\propto n^{1/d}$, which
is realized by 
atomic and molecular interaction by centrifugal or electric forces;
they practically interact only
with others nearby.
With the Pareto $f(w)$, which 
facilitates more global interaction,
we obtain
\begin{eqnarray}
k(w) &=& \rho
\int^{\infty}_{w_0}\frac{a}{w_0}\left(\frac{w_0}{w^{\prime}}
\right)^{a+1}
\pi^{d/2}\Gamma\left(\frac{d}{2}+1\right)
\left( \frac{w w^{\prime}}{\theta}\right)^{d/\beta}
dw^{\prime} \nonumber\\
&=& c_2 \; w^{d/\beta}, 
\label{eq:k(w)_gravity}
\end{eqnarray}
where 
\begin{equation}
c_2 = 
\frac{\rho \pi^{d/2}}{\theta^{d/\beta}}
\frac{a}{a-\frac{d}{\beta}} w_0^{d/\beta}
\; \Gamma\left(\frac{d}{2}+1 \right).
\label{eq:c_2}
\end{equation}
Equation~(\ref{eq:k(w)_gravity}) is essentially the same as
\EQ(\ref{eq:k(w)_pareto_power}), and $\beta>d/a$ must be satisfied 
for $c_2>0$. The original gravity model for social
interaction has $\beta=1$ and $d=2$
\cite{Zipf47}, and hence $a>2$ is necessary.
Combination of \EQS(\ref{eq:f(w)_pareto}) and (\ref{eq:k(w)_gravity})
yields
\begin{equation}
p(k) = \frac{a\beta w_0^a}{c_2 d} w^{-a-(d/\beta)}
= \frac{a\beta c_2^{a\beta/d} w_0^a}{d} 
k^{-1-(a\beta/d)}. 
\label{eq:p(k)_gravity}
\end{equation}
When $ r^{\beta} > w_0^2/ \theta$,
we obtain
\begin{eqnarray}
g(r) &=& \int^{\theta r^{\beta} / w_0}_{w_0}
\frac{a}{w_0}\left(\frac{w_0}{w}\right)^{a+1}
\left(\frac{w w_0}{\theta r^{\beta}} \right)^a dw 
+ \int^{\infty}_{\theta r^{\beta} / w_0}
\frac{a}{w_0}\left(\frac{w_0}{w}\right)^{a+1} dw\nonumber\\
&=& \frac{w_0^{2a}}{\theta^a} 
\left( a\log\frac{\theta r^{\beta}}{w_0^2} + 1 \right) r^{-a\beta}.
\label{eq:g(r)_gravity}
\end{eqnarray}
Comparison of \EQS(\ref{eq:p(k)_add_pareto}) and
(\ref{eq:g(r)_add_pareto}) with \EQS(\ref{eq:p(k)_gravity}) and
(\ref{eq:g(r)_gravity}) reveals that the asymptotic behavior
of $p(k)$ and that of $g(r)$ coincide with those of the additive
weight model with the Pareto $f(w)$ and
$h(r)=r^{-\beta}$. Given the Pareto $f(w)$ and $h(r)=r^{-\beta}$,
whether multiplicative or
additive interaction is used does not matter so much.

Numerically evaluated
$L$, $C$, $p(k)$, and $C(k)$ for varying $\beta$ are shown in
\FIG\ref{fig:gravity}. We set $n=10000$, $a=1$, $w_0=1$, and
\begin{equation}
\left< k\right> =
\int^{\infty}_{c_2 w_0^{d/\beta}} k\; p(k)\; dk
= \frac{a\beta c_2 w_0^{d/\beta}}{a\beta - d}
\label{eq:<k>_gravity}
\end{equation}
constant at 20. Figures~\ref{fig:gravity}(a) and \ref{fig:gravity}(b)
show that $L$ and $C$ have a similar dependence on $\beta$ to the
additive weight model with exponential $f(w)$
[\FIGS\ref{fig:exp_add}(a) and \ref{fig:exp_add}(b)].
Figure~\ref{fig:gravity}(a) indicates that a transition from a small-$L$
 regime to a large-$L$ regime occurs somewhere around $\beta=3$.
Since \FIG\ref{fig:gravity}(b) supports that $C$
remains finite for large $n$ irrespective of $\beta$, the small-world
property is suggested for small $\beta$.  The transition appears
similar to the phase transition in geographical BA models
\cite{Manna,Xulvi}.  However, in those models,
 $\gamma$ does not change in $\beta>0$ as
far as the network is in the small-world regime,
whereas it does change here (but see \cite{Rozenfeld}).  As shown in
\FIGS\ref{fig:gravity}(c), \ref{fig:gravity}(d), and \ref{fig:gravity}(e),
$p(k)$ (crosses) obey power laws whose
scaling exponents are well predicted by \EQ(\ref{eq:p(k)_gravity})
(lines).  Consequently, the weighted gravity model realizes scale-free
small-world networks when $\beta$ is small enough.
In this scheme, $\gamma$ is tunable by varying $a$, $\beta$, and $d$.
Circles in \FIGS\ref{fig:gravity}(c),
\ref{fig:gravity}(d), and \ref{fig:gravity}(e)
indicate $C(k)\propto k^{-\gamma^{\prime}}$ with
$\gamma^{\prime}\cong 2$ or somewhat smaller.
Finally, numerically obtained $g(r)$ shown by circles 
in \FIGS\ref{fig:gr_gravity}(a) ($\beta=2$) and
\ref{fig:gr_gravity}(b) ($\beta=3$) decays algebraically as
predicted by \EQ(\ref{eq:g(r)_gravity}).

A generated network is shown in
\FIG\ref{fig:wattslike} for $n=100$, $d=1$, $a=1$, $w_0=1$,
$\beta=1$, and hence $\gamma=2$. For demonstration purposes, the
vertices are aligned on a one-dimensional ring.  In spite of the small
size, the figure is indicative of the scale-free and small-world
properties. It is visually comparable to the BA-type scale-free small-world
networks on a ring \cite{Xulvi} and the Watts-Strogatz non-scale-free
small-world networks \cite{SW}.

In other geographical network models, $L$ becomes large if $g(r)$
decays faster than $g(r)\propto r^{-\delta}$ with a certain
$\delta>0$.  For example, a nonscale-free weightless network model on
a lattice owns an ultrasmall $L=O(1)$ for $\delta\le d$, small
$L=O(\log n)$ for $d< \delta < 2d$, and large $L=O(n^{1/d})$ for
$\delta\ge 2d$ \cite{Biskup03}. In another nonscale-free network,
$\delta=d+1$ divides the small-world and large-world regimes
\cite{Sen02}.  Also in a one-dimensional geographical scale-free
network model with preferential attachment, a similar phase transition
occurs at $\delta=1$ \cite{Xulvi,Sen03}. Based on
\FIGS~\ref{fig:exp_add}(a) and \ref{fig:gravity}(a), we anticipate
that the gravity model has the phase transition at a critical value
$\delta$ under which the network is geographical, scale-free, and
small-world at the same time. We do not examine $h(r)$ decaying faster
than $\left(\log r\right)^{-1}$ in the additive weight notation
[\EQ(\ref{eq:rule_h_log})] or equivalently $h(r)$ decaying faster than
algebraically in the multiplicative weight notation
[\EQ(\ref{eq:rule_gravity})], for which we expect too large $L$. Let
us mention that $h(r)\propto (\log r)^{-1}$, which other models have
largely neglected, may be appropriate if weight interaction is
effectively additive.

The results in \SEC\ref{sub:pareto_power} and those in 
this section can be
captured as a spatial extension of the results in 
\cite{Servedio}, which addresses the
inverse problem to determine $f(w)$ from $p(k)$.
To obtain $p(k)\propto k^{-\gamma}$, a
pair of vertices with weights $w$ and $w^{\prime}$ that
follow $f(w)=\lambda e^{-\lambda w}$ with $\lambda = 1$ are connected
with probability proportional to
$\exp\left[-(w+w^{\prime})/(-\gamma+1)\right]$ \cite{Servedio}.  In
the gravity model, we have defined $W=\exp(w)$ and
$W^{\prime}=\exp(w^{\prime})$ so that $W$ and $W^{\prime}$ follow the
Pareto distribution. The probability that the two vertices are
connected is proportional to the volume of a $d$-dimensional ball
with radius $r_0$, where $W W^{\prime}/r_0^{\beta} = \theta$.
This probability is proportional to $ r_0^d \propto \left( W
W^{\prime}/\theta\right)^{d/\beta} \propto
\exp[(w+w^{\prime})d/\beta]$. We should have $d/\beta = 1/(\gamma-1)$,
which is consistent with \EQ(\ref{eq:p(k)_gravity}) since
$a=\lambda=1$.

\subsection{Gravity model with exponential $f(w)$}

Let us treat the gravity model with
the exponential weight distribution. This configuration
is equivalent to the model with additive weight interaction,
$h(r)\propto (\log r)^{-1}$, and
a weight distribution less broad than the exponential
distribution. It follows that
\begin{equation}
k(w) = \rho \int^{\infty}_0 \lambda e^{-\lambda w^{\prime}}
 \pi^{d/2}\Gamma\left(\frac{d}{2}+1\right)
\left( \frac{w w^{\prime}}{\theta}\right)^{d/\beta} dw^{\prime}
= c_1 \Gamma\left(\frac{d}{2}+1\right) w^{d/\beta}
\end{equation}
and
\begin{equation}
p(k) = \frac{\beta\lambda}{d k^{1-\beta/d}
\left[c_1\Gamma\left(d/\beta+1\right) \right]^{\beta/d}}
\exp\left[-\lambda
\left(\frac{k}{c_1\Gamma\left(d/\beta+1\right)}\right)^{\beta/d}
\right],
\end{equation}
which is a stretched exponential with a modifying factor 
$k^{-1+\beta/d}$. Furthermore, we have 
\begin{eqnarray}
g(r) &=& \int^{\infty}_0 \lambda e^{-\lambda w} 
e^{-\lambda\theta r^{\beta}/w}dw
= 4\lambda^2 \int^{\infty}_0 \frac{\cos (\sqrt{\theta r^{\beta}}u)}
{(u^2+4\lambda^2)^{3/2}}du\nonumber\\
%
%
&=& 4\lambda^2 \int^{\infty}_1 \frac{e^{-2\lambda\sqrt{\theta
     r^{\beta}}t}}{\sqrt{t^2-1}} dt
= 4\lambda^2 K_0 (2\lambda \sqrt{\theta r^{\beta}}),
\label{eq:g(r)_gravity_exp}
\end{eqnarray}
where $K_0(x)$ is the modified Bessel function of the second kind
\cite[pp.185, 187--188, and 206]{Watsonbook}.
Since $K_0(x)$ tends to
\begin{equation}
K_0(x) = \sqrt{\frac{\pi}{2x}} e^{-x} \left[ 1 - \frac{1}{8x} +
O\left( \frac{1}{x^2}\right)\right]
\end{equation}
as $x\to\infty$ \cite[p.202]{Watsonbook}, 
\EQ(\ref{eq:g(r)_gravity_exp}) asymptotically behaves as
\begin{equation}
g(r) \cong 2\pi^{1/2}\lambda^{3/2}
\left(\theta r^{\beta}\right)^{-1/4}
e^{-2\lambda\sqrt{\theta r^{\beta}}}\quad (r\to\infty).
\label{eq:g(r)_gravity_exp_final}
\end{equation}
With the arguments in \SEC\ref{sec:gravity} taken into account,
\EQ(\ref{eq:g(r)_gravity_exp_final}) implies that $g(r)$ decays too
fast to make the network small-world.  A lesson is that $f(w)$
considerably influences network properties, which is not the case for the
nongeographical counterpart \cite{Masuda_THRESH}.  Particularly, the
Pareto $f(w)$ can yield scale-free $p(k)$ and the small-world properties,
regardless of whether weight interaction is additive or multiplicative.
On the other hand, the exponential $f(w)$ explored in this section
and \SEC\ref{sub:exponential_power} induces exponential types of $p(k)$
and large $L$.

\section{Scale-free networks and scale-free weight distributions}\label{sec:discussion}

Among the configurations considered in \SEC\ref{sec:model}, the
additive weight model and the gravity model with scale-free $f(w)$ and
scale-free $h(r)$ generate small-world networks with scale-free
$p(k)$.  In this
regime, our model relates scale-free $p(k)$, which is of
recent research interest, to general power law distributions in
nature that have a long history tracing back to Pareto.
Let us discuss the relevance of our model to real data.

There is a body of evidence that quantities potentially serving as
vertex weights are distributed according to power laws $f(w)\propto
w^{-a-1}$.  For example, the celebrated Pareto and Zipf laws dictate
that incomes and city sizes follow power laws with $a+1=2.0$
\cite{Zipf47,Zipflaw}.  More recent data
analyses confirm power laws in
countrywise city sizes ($a+1=1.81$--$2.96$ with mean $a+1=2.136$)
\cite{Rosen}, firm sizes ($a+1=2$) \cite{Axtell}, the number of pages per
website ($a+1=1.65$--$1.91$) \cite{Huberman}, land prices
($a+1=2.1$--$2.76$) \cite{Kaizoji_Andersson}, incomes
($a+1=1.7$--$2.4$) \cite{Levy_Okuyama}, and importance of
airports ($a+1 = 1.67$) \cite{Barrat}, to name a few.  On the other
hand, the original gravity model disregarding weight distributions
assumes $\beta=1$ and $d=2$ \cite{Zipf47}. Application of the
values of $a$ mentioned above to the weighted gravity model yields
$\gamma =1+a\beta/d = 1.32$--$1.98$, which is
too small to fit real network data whose $\gamma$ mostly
falls between 2 and 3 \cite{reviews}.  As another indication, an
extensive data analysis of the Internet revealed $g(r)\propto
r^{-\delta}$ with $\delta=1$ \cite{Yook02}.  If our model could
underlie the Internet, it should mean $a\beta=\delta=1$, and hence
$\gamma = 1 + a\beta/d = 3/2$ since $d=2$.  This $\gamma$ is again too
small for the real Internet and related computer networks that have
$\gamma=1.9$--$2.8$ \cite{reviews}.

However, we regard that our model is not necessarily implausible.
First, our model and also the nonspatial
threshold model do not aim to describe growing networks;
the Internet is a typical example of
growing networks \cite{reviews}.
Our goal is rather to discuss nongrowing networks in a
geographical context.  As a supporting example, connectivity networks
of brain regions have $\gamma=2$, $\delta=a\beta\cong 2$, and
$d=2$ \cite{Eguiluz03}, which are roughly consistent with
\EQ(\ref{eq:p(k)_gravity}). Actually, the brain network does not
grow so much once an animal is born.

Second, estimation of $\beta$ involves much fluctuation because of
the difficulty in data acquisition. Since the proposal of
the gravity model in which $\beta=1$ was inferred from railway and
highway travel data \cite{Zipf47}, analyses of various social
activity data have offered a wide range of $\beta$.  Among them are
investigations of air travels ($\beta=0.2$--$2.0$)
\cite{Haynes,Long_Howrey,Fotheringham}, journey to work
($\beta=0.5$--$1.2$) \cite{Hutchinson}, migration ($\beta= 1.59,
2.49$) \cite{Morrill}, cedar rapids direct contacts ($\beta=2.74$)
\cite{Morrill}, marriage ($\beta=0.59, 1.53, 1.59$) \cite{Morrill},
 and memorizable
social interaction ($\beta=2$) \cite{Latane}.
The ambiguity and the activity dependence of $\beta$ render
the evaluation of $\gamma$ pretty uncertain.  Precision of $\beta$ in
classical studies was also low because of small data sizes. To
undertake more detailed and large-scale data analysis as in
\cite{Yook02,Latane} is important.

Third, the interaction strength, which is assumed to be proportional
to $w_1 w_2/r^{\beta}$ in the gravity model, may be nonlinear in
weights.  For example, use of $w_1^x
w_2^x/r^{\beta}$ \cite{Haynes} results in
$\gamma=1+a\beta/xd$. Real data actually support
$0.73\le x\le 1.05$ \cite{Long_Howrey}, and $x$
smaller than 1 increases $\gamma$ to make it more realistic.  By the
same token, replacing $(w_1+w_2)/r^{\beta}$ with
$(w_1+w_2)^x/r^{\beta}$ in the additive notation effectively changes
$\beta$ to $\beta/x$.

Next, let us relate our model to network search problems in which an
agent on a vertex attempts to reach an unknown destination by
traveling on edges. In small-world networks 
defined by
lattices supplied with long-range connections with density $g(r)\propto
r^{-\delta}$, which are essentially equivalent to
the random connection model \cite{Meester_Roy},
optimal search performance is realized when $\delta=d$
\cite{Kleinberg_Franceschetti}.  Even though the weighted gravity
model is a different model, simple adoption of
our formula suggests $d = \delta = a\beta$ and $\gamma = 1+a\beta/d=2$.
Computer-related networks usually have $\gamma>2$ presumably because
they are growing.  However, some social networks and peer-to-peer
networks, which may be considered to be nongrowing, own $\gamma$ close
to 2 \cite{reviews}, enhancing the search ability.

Similarly, emergence of small-world networks in a geographical
framework requires $d+1 > \delta > d$, while latticelike networks
result from $\delta>d+1$, and $\delta<d$ induces randomlike networks
with low clustering \cite{Sen02}. Simple-minded substitution of
$\delta=a\beta$ leads to $d+1 > a\beta > d$ and
$2<\gamma<2+ d^{-1}$. Since we usually have $d=2$ or $3$,
$\gamma$ associated with general nongrowing
small-world networks may be close to 2.  To summarize,
scale-free networks with $\gamma$ around 2 may be optimal in the sense
of the search performance and the small-world property.  In addition,
$\gamma=2$ is the baseline scaling exponent of the nongeographical 
threshold graph
\cite{Masuda_THRESH}, and it may also be the case for general
cooperative nongrowing networks
\cite{Caldarelli,Goh_Chung,Toroczkai,Boguna,Servedio,Masuda_THRESH,Rozenfeld}.
In contrast, $\gamma=3$ is an important phase-transition point for
percolation and dynamic epidemic processes \cite{cohen_pastor}.  The BA
model has $\gamma=3$, which may set the baseline $\gamma$
for other competitive
growing networks
\cite{reviews,Bianconi,Manna,Xulvi,Yook02,Sen03,Ravasz}.
Our current speculation stems from the ansatz
$\gamma=1+a\beta/d = 1 + \delta/d$ plugged 
into the results obtained from other models.
Further investigation of this issue is an important future problem.

\section{Conclusions}

We have proposed and analyzed 
a geographical nongrowing network model based on
thresholding the sum of two vertex weights. Our model 
contrasts with geographical growing models based on the BA model, and
it naturally extends
the threshold graph, the unit disk graph, and the gravity model,
which are widely used in a range of fields.  In proper regimes,
small-world networks with scale-free degree distributions and the
connection probability algebraically decaying in distance are
generated, and they are consistent with many real data.  In contrast
to the nongeographical threshold model, what weight
distribution is used 
matters for network properties. For scale-free networks
to emerge, the weight should be distributed as specified by power
laws.  The weight
distribution and the degree distribution generally have different
scaling exponents, and they are bridged by a relation
involving the spatial dimension and the decay rate of the interaction
strength.

\begin{acknowledgements}
We thank D. Harada for assistance in the numerical simulations in this
work and S. Havlin and Y. Otake for helpful comments related to
this work.  This study is supported by
RIKEN and a Grant-in-Aid for Young Scientists (B) (Grant No. 15700020)
of Japan Society of the Promotion of Science.
\end{acknowledgements}

\newpage

Figure captions

\bigskip

Figure 1: Network properties with $h(r)=r^{-\beta}$ and the
exponential weight distribution with $\lambda = 1$ and $\left<k\right>=20$.
Dependence of (a) $L$ and (b) $C$ on $\beta$ for
$n=2000$ (thinnest lines),
4000, 6000, 8000, and 10000 (thickest lines) is presented.
The relation between $L$ and $n$ is shown in 
the inset of (a), with
upper lines corresponding to larger $\beta$.
Also shown are numerically obtained
$p(k)$ (crosses) and $C(k)$ (circles) with $n=10000$ for
(c) $\beta=0.5$, (d) $\beta=1.5$, and (e) $\beta= 2.5$.

\bigskip

Figure 2: Network properties for the gravity model with the Pareto
weight distribution with $a=1$, $w_0=1$, and $\left<k\right>=20$.
Dependence of (a) $L$ and (b) $C$ on $\beta$ is presented. 
Also shown are numerically obtained
$p(k)$ (crosses), $C(k)$ (circles) with $n=10000$, and 
the theoretical prediction $p(k)\propto k^{-1-a\beta/d}$ (lines)
for (c) $\beta=2$, (d) $\beta=3$, and (e) $\beta= 4$.

\bigskip

\begin{figure}
\begin{center}
\includegraphics[height=2.25in,width=2.25in]{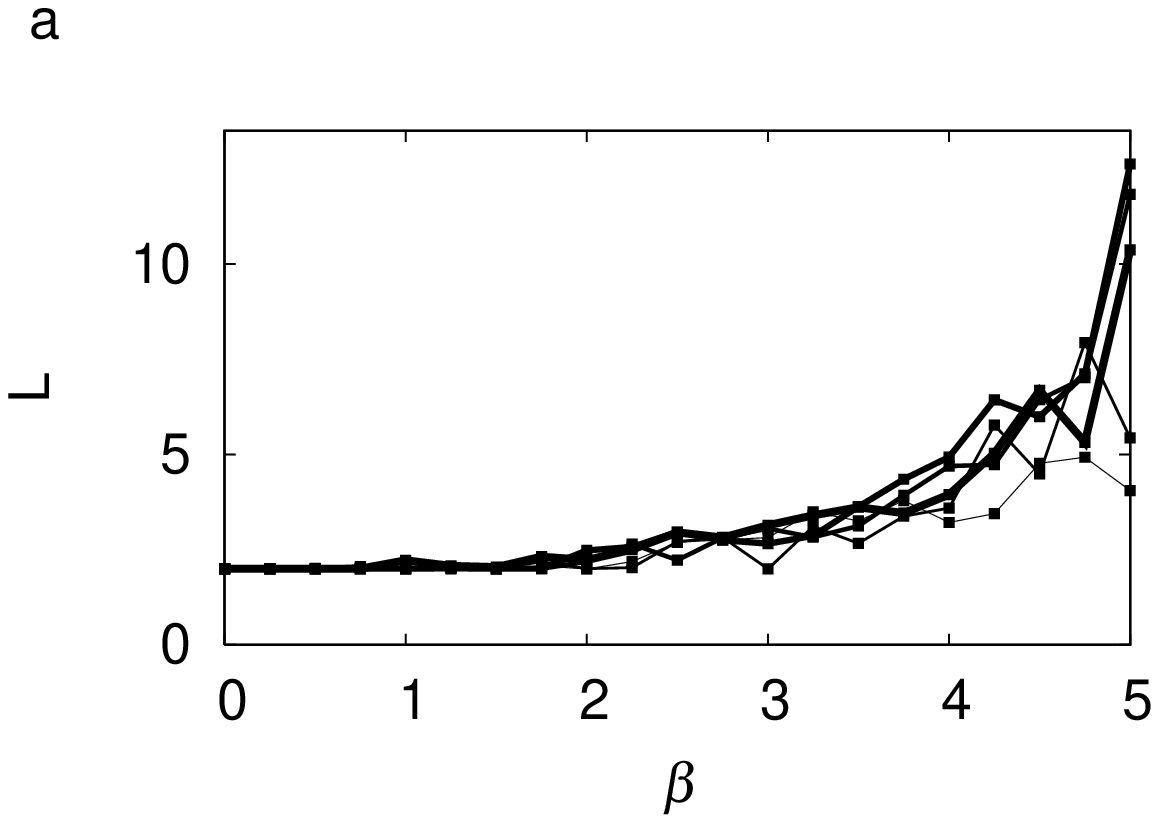}
\includegraphics[height=2.25in,width=2.25in]{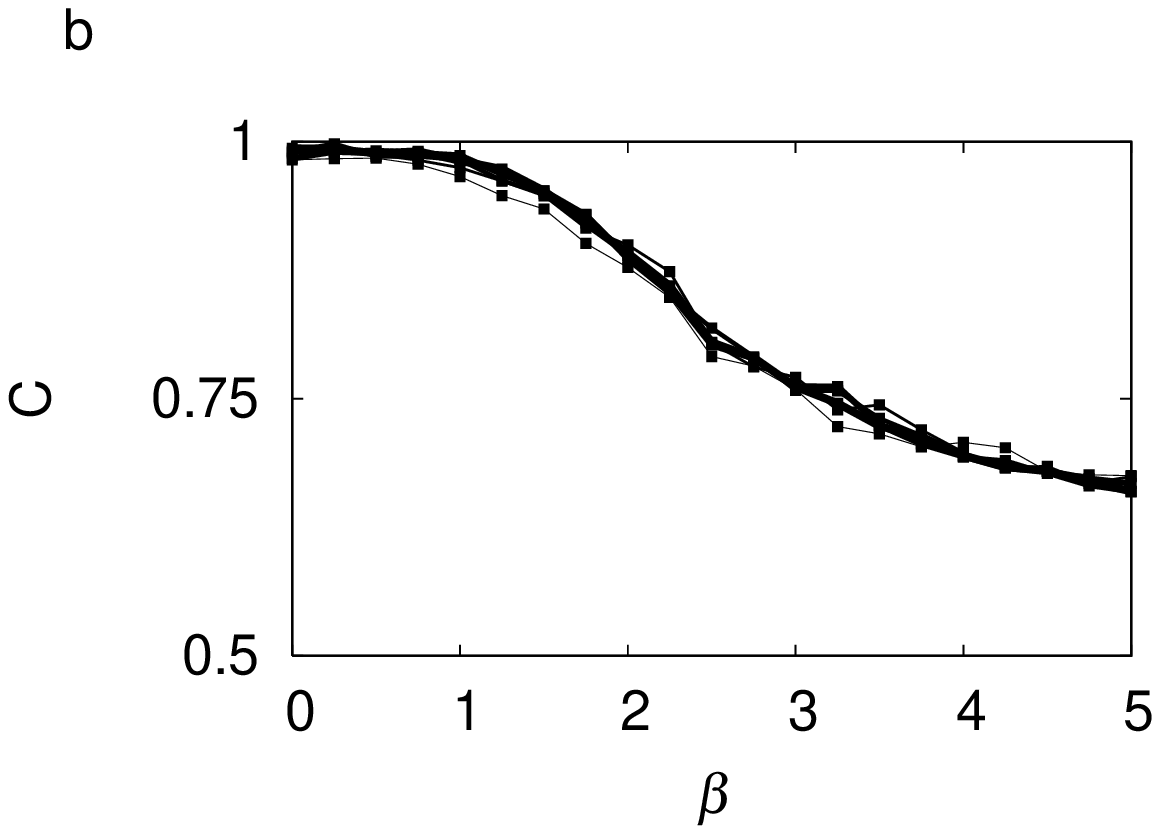}
\includegraphics[height=2.25in,width=2.75in]{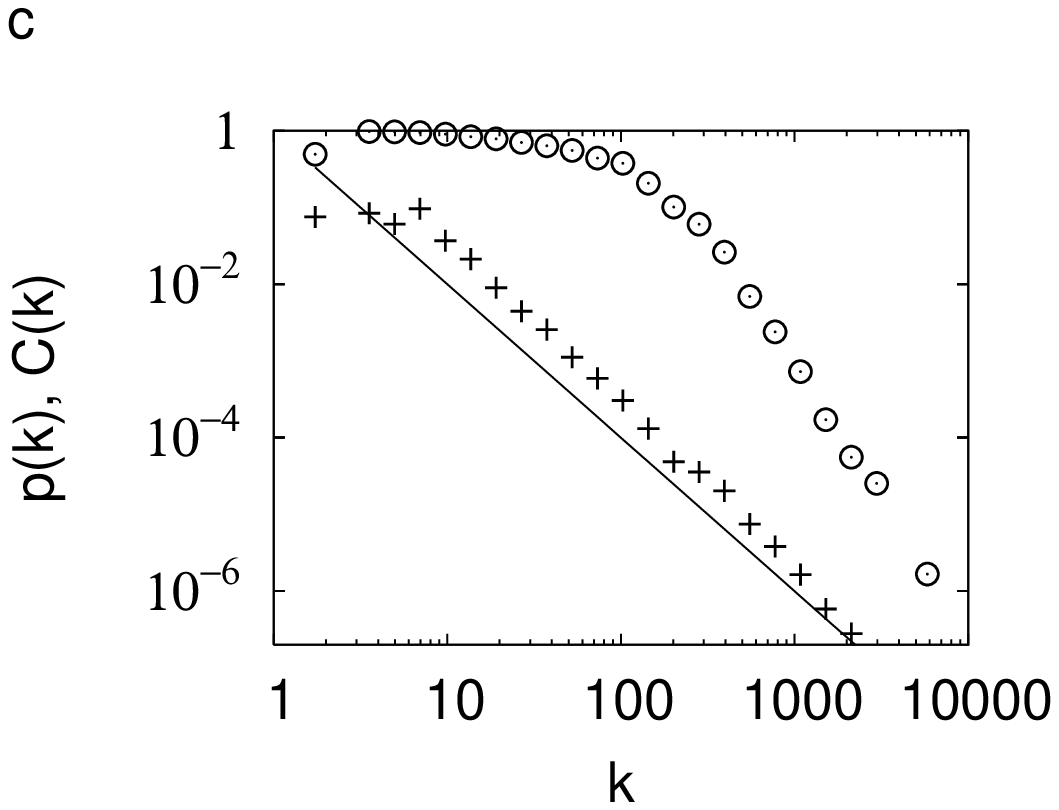}
\includegraphics[height=2.25in,width=2.75in]{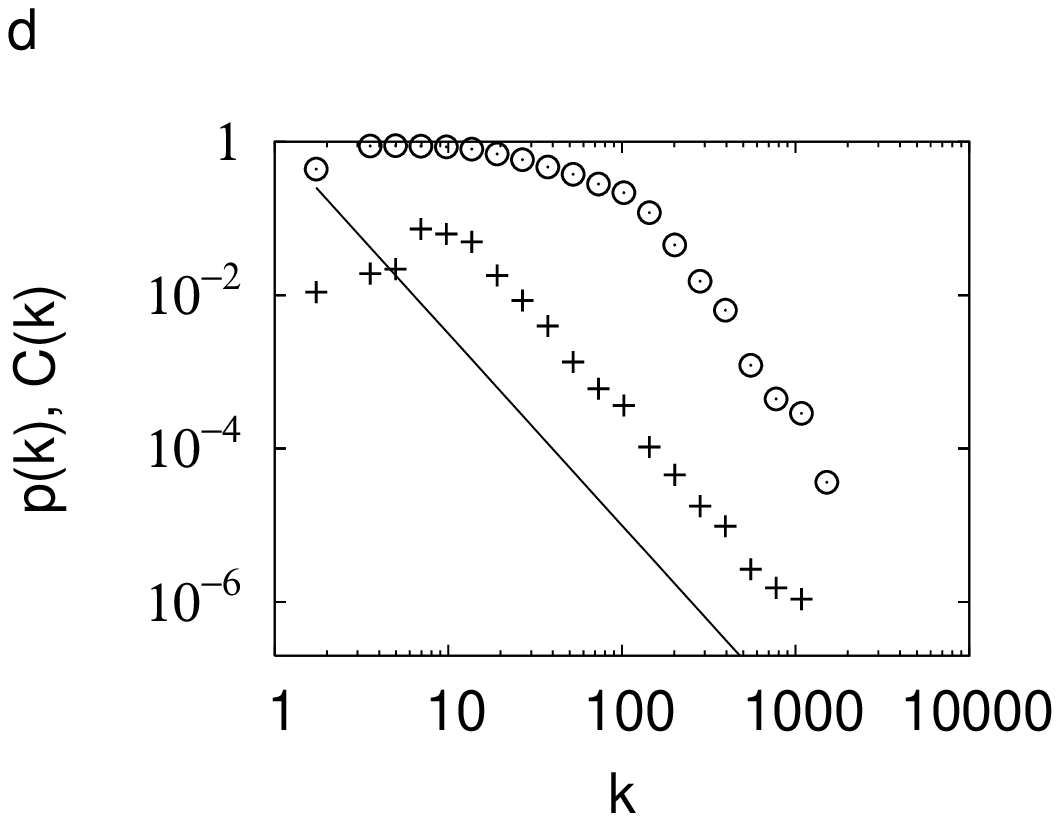}
\includegraphics[height=2.25in,width=2.75in]{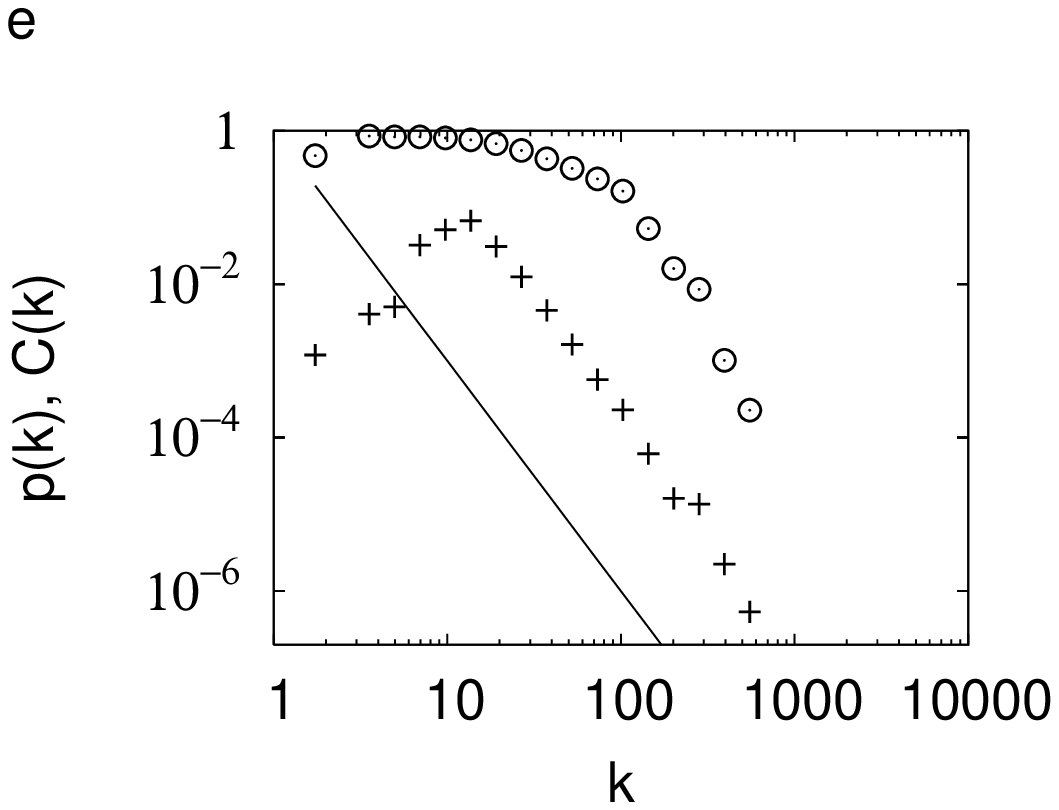}
\caption{}
\label{fig:gravity}
\end{center}
\end{figure}

\begin{figure}
\begin{center}
\includegraphics[height=2.25in,width=2.75in]{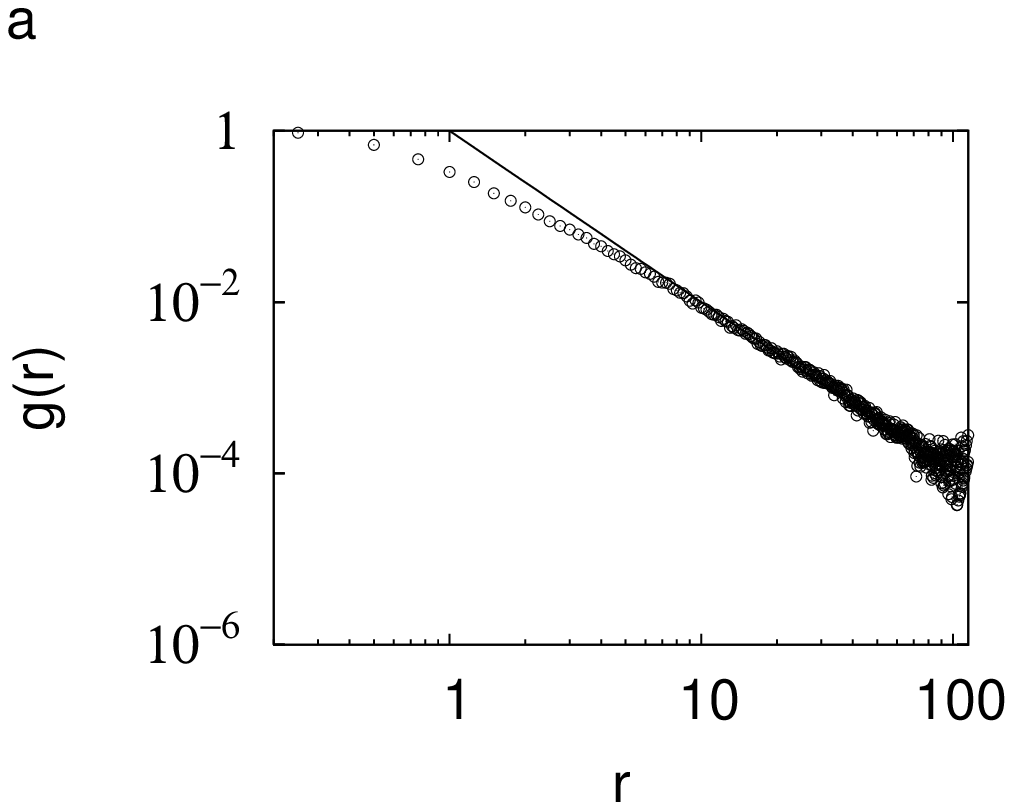}
\includegraphics[height=2.25in,width=2.75in]{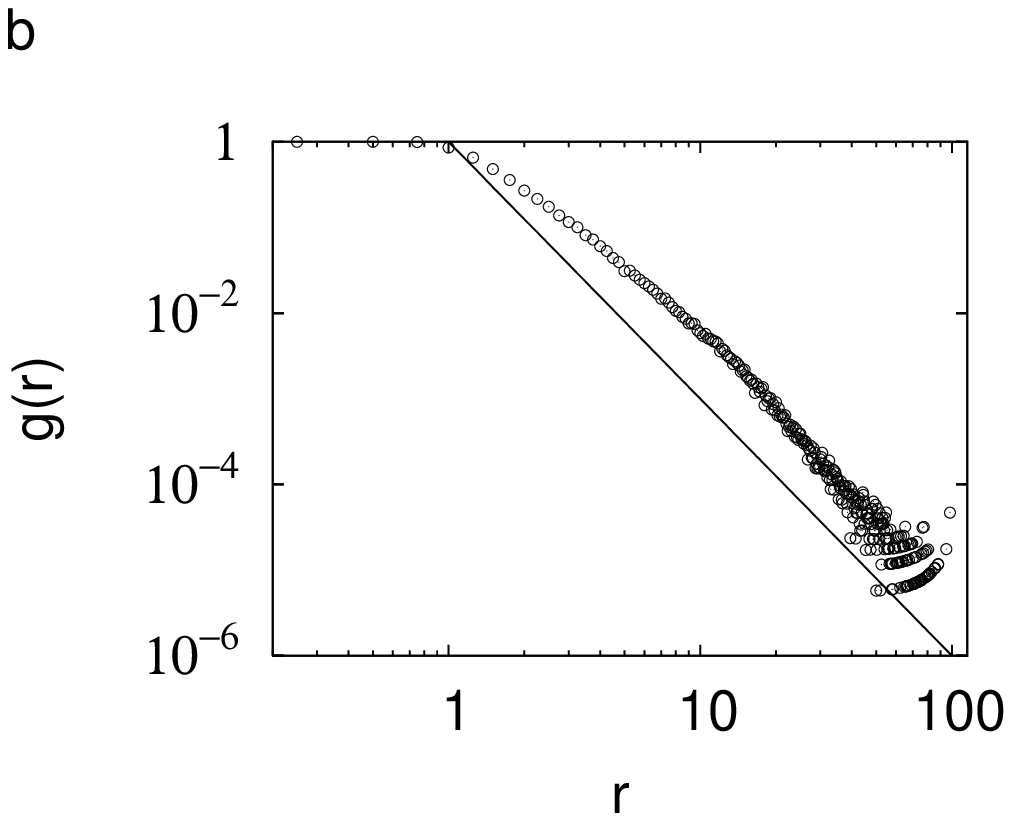}
\caption{}
\label{fig:gr_gravity}
\end{center}
\end{figure}

\begin{figure}
\begin{center}
\includegraphics[height=3.25in,width=3.25in]{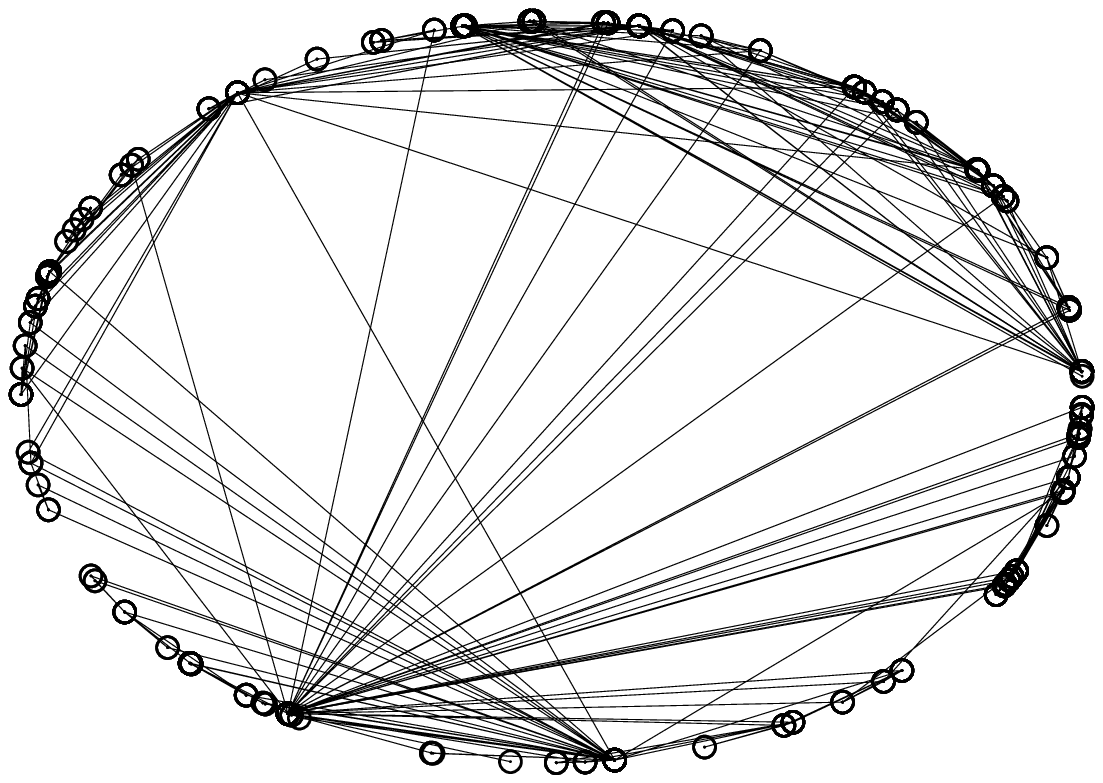}
\caption{}
\label{fig:wattslike}
\end{center}
\end{figure}

Figure 3: Numerically obtained $g(r)$
(circles) and the prediction by \EQ(\ref{eq:g(r)_gravity}) (lines)
for the gravity model with (a) $\beta=2$ and (b) $\beta=3$. 
The other parameter values are the same as those used in
\FIG\ref{fig:gravity}.

\bigskip

Figure 4: An example of the weighted gravity model
on a one-dimensional ring. We set $n=100$, $\beta=1$,
and $\left<k\right>=6$.
The Pareto weight distribution with $a=1$ and $w_0=1$ is used.

\newpage
\clearpage

\clearpage

\clearpage

\clearpage

\end{document}